\documentclass[%
aps,pre,
twocolumn,
 groupedaddress,showpacs,showkeys
]{revtex4-1} 

\usepackage{amsmath,amssymb}
\usepackage{mathtools}
\usepackage{graphicx}% Include figure files
\usepackage{bm}% bold math
\usepackage{color}
\usepackage{subfigure}
\renewcommand{\underline}[1]{#1}
\newcommand{\bunderline}[1]{#1} %{\underline{#1\mkern-4mu}\mkern4mu}

\newcommand{\W}{\bunderline{\mathbf{B}}}
\newcommand{\Ws}{B}
\newcommand{\uu}{\underline{\mathbf{u}}}
\newcommand{\uvarphi}{\underline{\boldsymbol\varphi}}
\newcommand{\uv}{\underline{\mathbf{v}}}
\newcommand{\uA}{\underline{\mathbf{A}}}

\newcommand{\ux}{\underline{\mathbf{x}}}
\newcommand{\uy}{\underline{\mathbf{y}}}
\newcommand{\HS}[1]{\stackrel{\circ}{H}\phantom{\hspace{-0.1mm}}^{1}_{#1}}
\newcommand{\HSd}[1]{{H}^{-1}_{#1^{-1}}}
\newcommand{\z}{\mathbf{z}}
\newcommand{\q}{\mathbf{q}}
\newcommand{\J}{\mathbf{J}}
\newcommand{\X}{\mathbf{X}}
\newcommand{\LL}{\mathbf{L}}
\newcommand{\RR}{\mathbf{R}}
\begin{document}

\preprint{AIP/123-QED}

\title[]{Entropy production and the geometry of dissipative evolution equations}

\author{Celia Reina$^1$ and Johannes Zimmer$^2$}
\email[]{creina@seas.upenn.edu, zimmer@maths.bath.ac.uk}
\address{$^1$Department of Mechanical Engineering and Applied Mechanics, \\ University of Pennsylvania, Philadelphia, PA 19104, USA}
\address{$^2$Department of Mathematical Sciences, University of Bath, Claverton Down, Bath BA2 7AY, UK}

\date{\today}

\begin{abstract} Purely dissipative evolution equations are often cast as gradient flow structures,
  $\dot{\z}=K(\z)DS(\z)$, where the variable $\z$ of interest evolves towards the maximum of a functional $S$ according
  to a metric defined by an operator $K$. While the functional often follows immediately from physical considerations
  (e.g., the thermodynamic entropy), the operator $K$ and the associated geometry does not necessarily so (e.g., Wasserstein geometry for
  diffusion). In this paper, we present a variational statement in the sense of maximum entropy production that
  directly delivers a relationship between the operator $K$ and the constraints of the system. In particular, the
  Wasserstein metric naturally arises here from the conservation of mass or energy, and depends on the Onsager
  resistivity tensor, which, itself, may be understood as another metric, as in the Steepest Entropy Ascent
  formalism. This new variational principle is exemplified here for the simultaneous evolution of conserved and
  non-conserved quantities in open systems. It thus extends the classical Onsager flux-force relationships and the
  associated variational statement to variables that do not have a flux associated to them. We further show that the
  metric structure $K$ is intimately linked to the celebrated Freidlin-Wentzell theory of stochastically perturbed
  gradient flows, and that the proposed variational principle encloses an infinite-dimensional fluctuation-dissipation
  statement.

\end{abstract}


\maketitle
Dissipative evolution equations (e.g., heat conduction, mass diffusion, interface motion) often follow variational principles, such as Onsager's least dissipation of energy~\cite{onsager1931reciprocal_I,onsager1931reciprocal_II} and extensions, in particular those based on maximum entropy production (MEPPs \cite{Martyushev2006,Dewar2013}) or Steepest Entropy Ascent (SEA)~\cite{beretta1987steepest,Beretta2014,Beretta2014b}). Mathematically, these equations are often of gradient flow type, that is, they can be described by the steepest ascent/descent of a functional, such as the entropy. Here descent has to be measured in a metric, which is neither provided by the aforementioned variational approaches, nor it is always intuitive (e.g., Wasserstein metric for diffusion processes). In this article, we establish a variational framework based on the
\emph{ansatz} of maximal entropy production which sheds light on the geometry of purely dissipative
evolution equations.  This new approach (1) delivers a construction of the gradient flow metric from conservation
constraints in the variational formulation; (2) extends Onsager's principle to simultaneously account for conserved and
non-conserved quantities in open systems; and (3) encloses an infinite-dimensional fluctuation-dissipation statement,
as shown from a large deviation argument for stochastically perturbed gradient flows. The diagram of
Fig.~\ref{Fig:Diagram} summarizes the connections established in this paper.

\section{Background}
We sketch some of the most closely related variational principles and provide a short summary on gradient flows. The body of literature, both classic and recent, on these two topics is too large to be reviewed comprehensively here. 

\subsection{Entropy production}

\begin{figure*}[t]
\begin{center}
    {\includegraphics[width=0.65\textwidth]{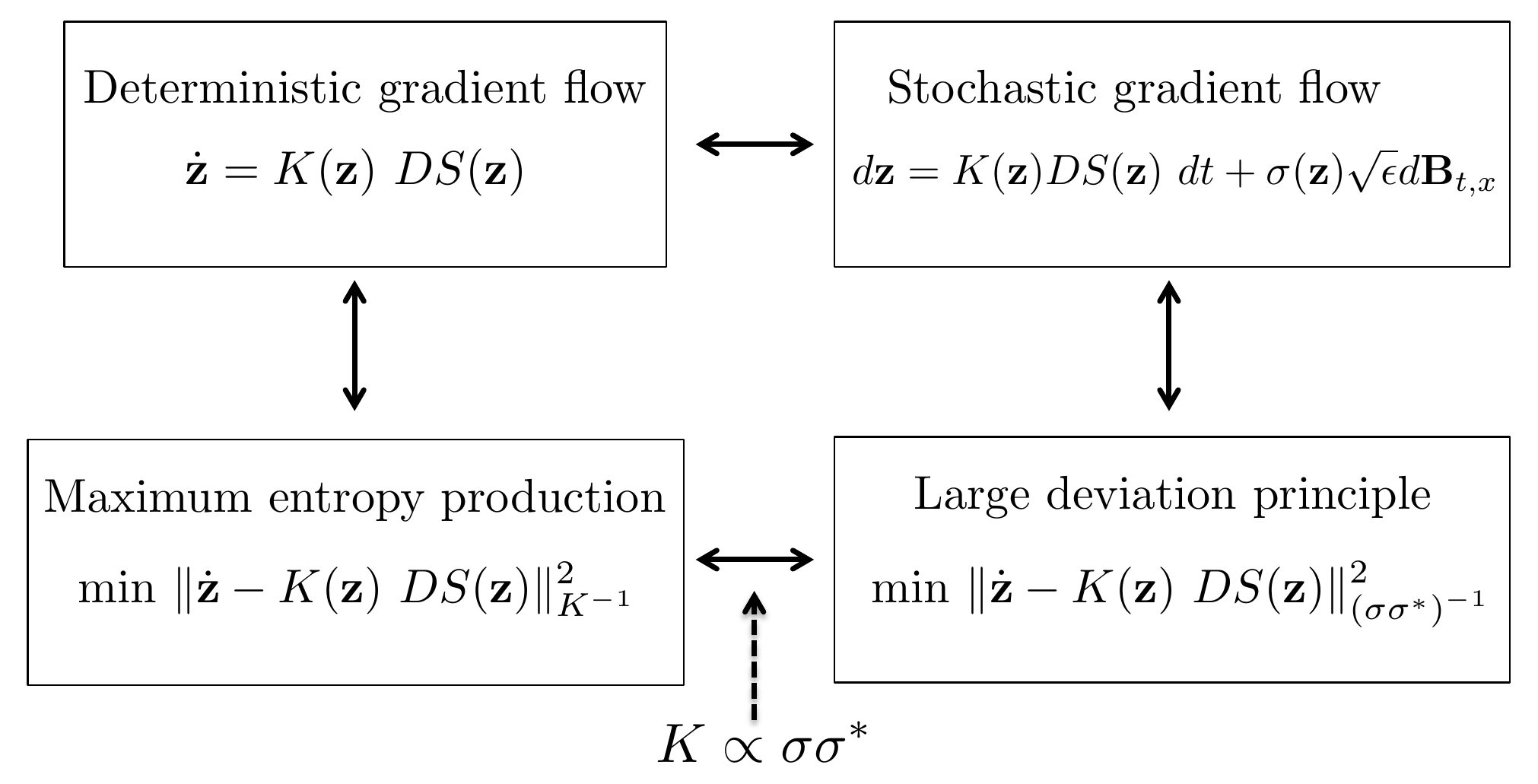}}
    \caption{Summary of the connections established in this work between the maximum entropy production principle, gradient flow structure, large deviation principle and fluctuation dissipation relation $K\propto\sigma \sigma^{*} $, with $\sigma$ defined by the stochastic gradient flow. }
    \label{Fig:Diagram}
\end{center}
\end{figure*}

Onsager, in his celebrated papers~\cite{onsager1931reciprocal_I,onsager1931reciprocal_II} generalized the transport laws, such as those by Fourier, Ohm or Fick, to account for a possible coupling between different physical processes. He proposed a general linear kinematic constitutive relation between fluxes $\J$ and forces $\X$, that is, $J_i = \sum_j L_{ij} X_j$. The conductivity matrix $\LL$ may depend on the state variables (temperature, pressure, chemical potential, etc.), but not on their gradient~\cite{gyarmati1970non}, and is symmetric as a result of the time reversal of the underlying atomistic equations of motion, $L_{ij}=L_{ji}$. These two properties of the constitutive relations --- linearity and symmetry of the conductivity tensor --- can be equivalently expressed by means of the \emph{principle of least dissipation of energy}~\cite{onsager1931reciprocal_I} (following Rayleigh's nomenclature~\cite{Rayleigh1913}). Namely, let $\sigma_s= \sum_iJ_iX_i$ be the entropy production and $\Phi(\J)=\frac{1}{2} \sum_{i,j}R_{ij} J_i J_j$ denote a local dissipation potential, with the resistivity tensor $\RR=\LL^{-1}$ being positive definite, then the variational principle reads 
\begin{equation} 
\label{Eq:LeastDissipationEngergy}
\max_{\J} [ \sigma_s(\J,\X) - \Phi(\J) ].
\end{equation}
In Onsager's words~\cite{onsager1931reciprocal_I}, `the rate of increase of the entropy plays the role of a
potential'. Several generalizations of this extremum principle have since emerged in different fields encompassing
climate \cite{Paltridge1975}, soft matter physics~\cite{doi2011onsager}, plasticity \cite{ziegler1983introduction}, biology \cite{Dewar2010} and quantum mechanics
\cite{Beretta1981} among others, and appear under the names of \emph{Maximum Entropy Production Principles}(MEPPs)
\cite{Martyushev2006,Dewar2013} and \emph{Steepest Entropy Ascent} (SEA)
\cite{beretta1987steepest,Beretta2014,Beretta2014b}. This latter framework provides a geometric
  interpretation of the resistivity tensor $\RR$ and generalizes $\Phi$ to arbitrary (but a priori unknown) metric spaces. Another approach to nonequilibrium thermodynamics, which combines reversible and irreversible dynamics, is the General Equation for NonEquilibrium Reversible-Irreversible Coupling (GENERIC)~\cite{grmela1997dynamics,ottinger2005beyond}. The structure of this formalism can be derived using contact forms in the setting of the Gibbs-Legendre manifold~\cite{grmela2014contact,grmela2015geometry}; it can be cast variationally; and it allows for a systematic multiscale approach~\cite{grmela2015reductions} as well as a treatment of fluctuations~\cite{grmela2014contact,grmela2012fluctuations}.

\begin{figure*} [t]
\begin{center}  
\subfigure[]{   
\includegraphics[width=0.8\columnwidth]{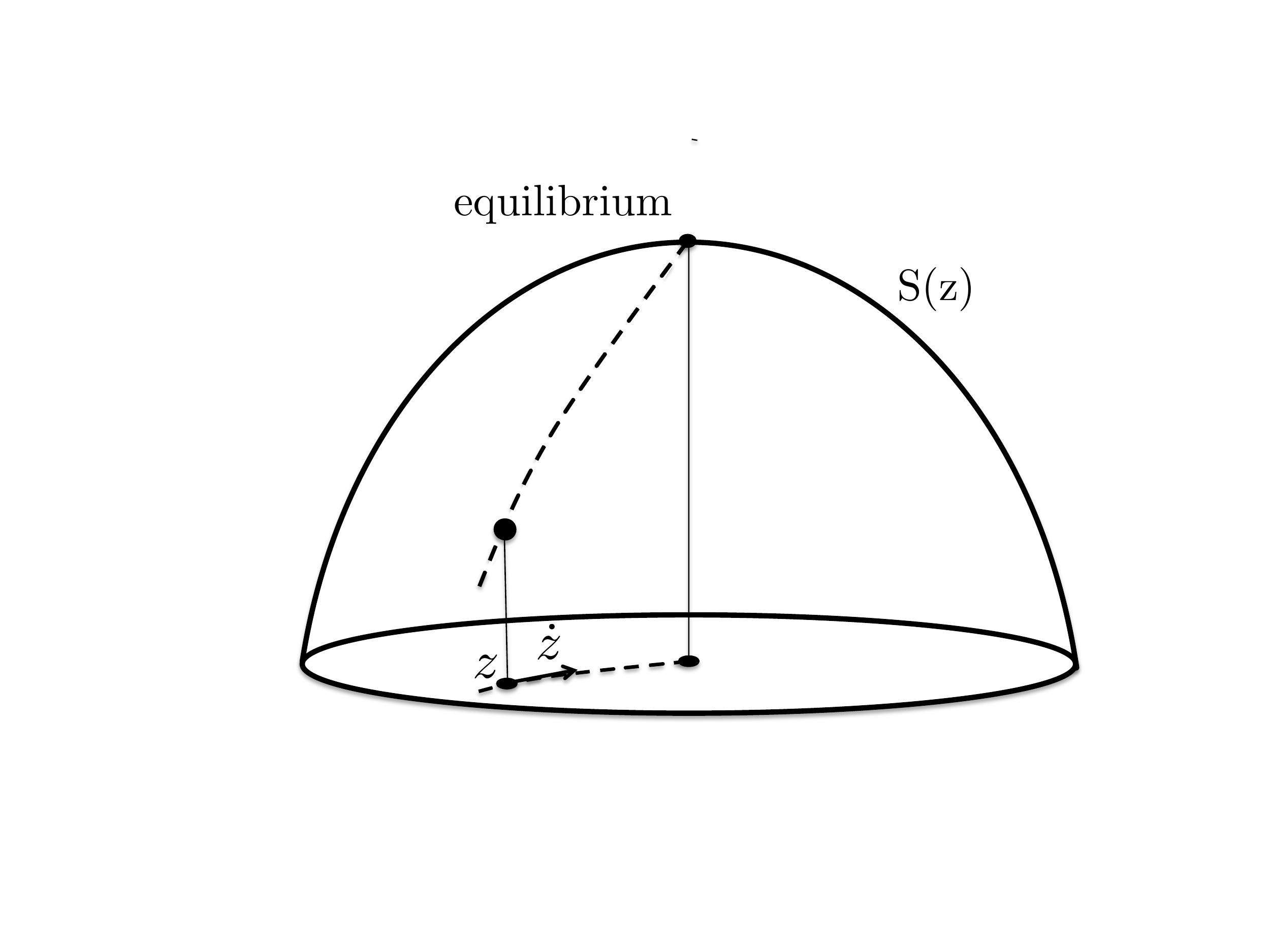} 
}    
\subfigure[]{     
\includegraphics[width=0.8\columnwidth]{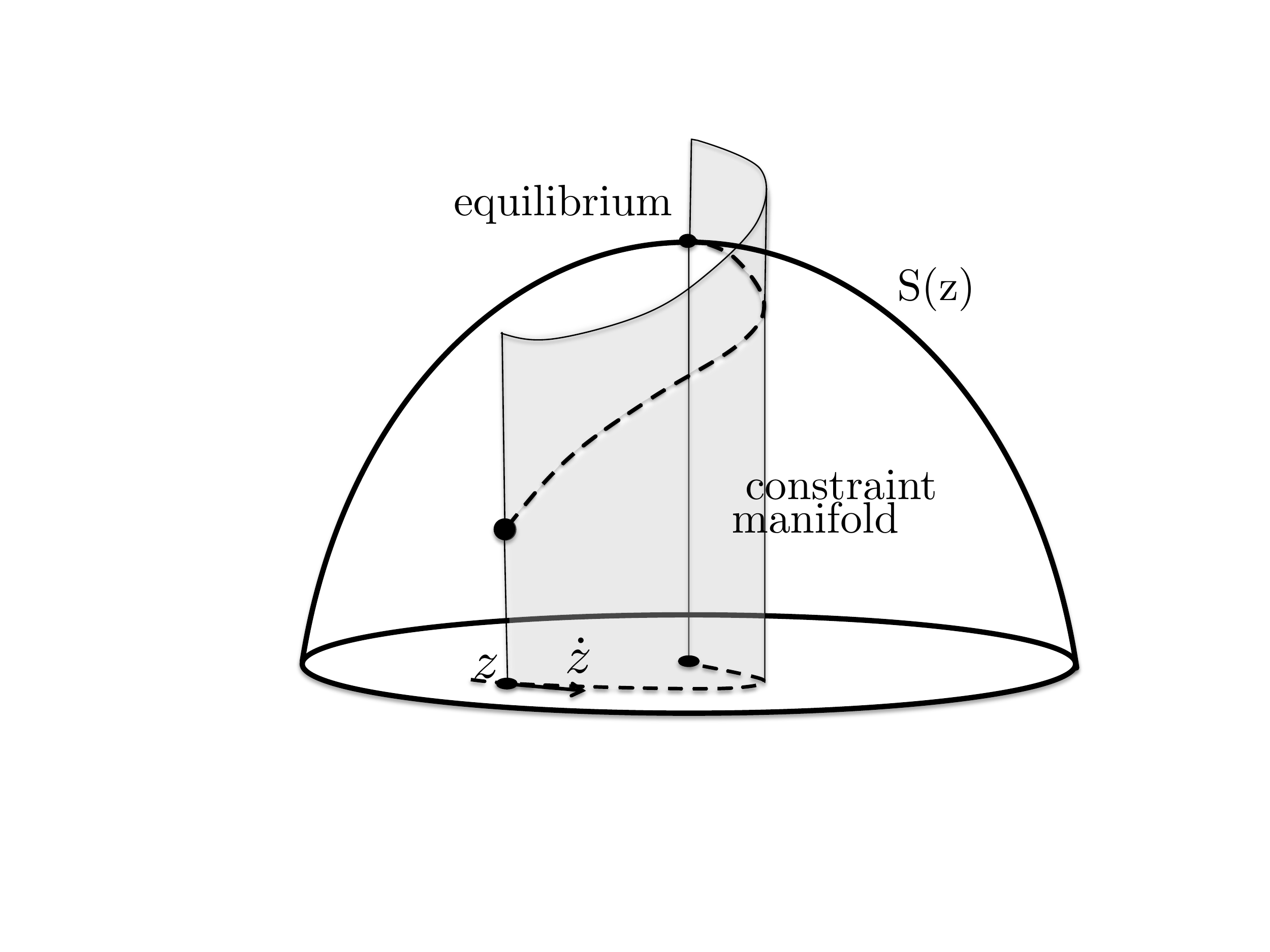}    
}    
\caption{Evolution of a closed system at $z(t)$ (a) without and (b) with the presence of a constraint. The equilibrium configuration coincides with the maximum of the entropy functional $S(z)$.} \label{Fig:ConstrainedMaximization}    
\end{center}
\end{figure*}

\subsection{Gradient flow structures}  
From a mathematical perspective, purely dissipative evolution equations can often be described as 
%Mathematicians have investigated dissipative evolution equations within the framework of 
\emph{gradient flow structures} \cite{ambrosio2006gradient}. %, where 
This means that the  vectorial variable $\z$  of interest (components are, for example, energy, density or interface position) evolves according to the steepest ascent of a functional $S$ (or descent for $-S$) in a geometry given by a metric associated with a positive semi-definite operator $K$,
\begin{equation}
\label{eq:abs-flow}
\dot{\z} = K(\z) DS(\z),\text { or }G(\z) \dot{\z} = DS(\z), 
\end{equation}
where $G=K^{-1}$ if the inverse is defined, and $DS\coloneqq \frac{\delta S}{\delta \z}$ is a force.  Note
that~\eqref{eq:abs-flow} is precisely the irreversible component of GENERIC. %(General Equation for NonEquilibrium Reversible-Irreversible Coupling) framework \cite{ottinger2005beyond}.  
Then $S$ is a Lyapunov functional,
$\dot{S} = \langle DS, \dot{\z} \rangle = \langle G\dot{\z}, \dot{\z} \rangle = \langle DS, K DS \rangle \geq 0$, where
$\langle,\rangle$ denotes the dual parity between elements of the tangent and the cotangent space.

Two common examples of \eqref{eq:abs-flow} are the $L^2_m$ flow and Wasserstein flow~\cite{JKO1998variational}, written for scalar-valued $z$ as
\begin{align}\label{Eq:L2_GF}
  &\dot{z} = m(z) DS(z), \\ \label{eq:wass}
  &\dot{z} = - \nabla \cdot \left( \mathbf{M}(z) \nabla DS(z) \right),
\end{align}
respectively, with $m \geq 0$ and $\mathbf{M}$ positive semi-definite. The latter equation is symbolically expressed as $\dot{z} = W_{\mathbf{M}}^{-1} DS(z), \text{or } W_{\mathbf{M}} \dot{z} = DS(z)$, with  $K \coloneqq W_{\mathbf{M}}^{-1} \coloneqq - \nabla \cdot \mathbf{M} \nabla $. Further details on the weak formulations of both flows and the norms involved are given in the Appendix.
%, detailed in the Appendix

It is noteworthy that the same equation can have different gradient flow representations. For example, the diffusion equation 
\begin{equation}
  \label{Eq:DiffusionEquation}
  \dot \rho(t,\ux) = \nabla \cdot (m(\rho(t,\ux)) \nabla \rho(t,\ux))
\end{equation}
can be interpreted both as $L^2_{m_1}$ flow (with mobility $m_1 \coloneqq m$ and Dirichlet integral $S_1(\rho) \coloneqq -\frac 1 2 \int |\nabla \rho|^2 \, dx$) and Wasserstein flow (with mobility $\mathbf{M}_2 \coloneqq m \rho\mathbf{I}$ and Boltzmann entropy $S_2=-\int_{\Omega} \rho \log \rho \, dx$).  The Wasserstein formulation is a natural choice since it involves the physical entropy. This flow and its associated metric will be automatically singled out by the variational principle proposed here, as we show next. %As will be shown in the following section, the Wasserstein metric will be automatically singled out by the newly proposed variational principle.
%It will be singled out by the newly proposed variational principle. 
%It is also useful for studying convergence to equilibrium~\cite{CarrilloToscani2004a}, and has natural generalizations for example to diffusion in porous media~\cite{otto2001geometry}. 

\section{Entropy production and deterministic evolution}\label{Sec:MaxEnt}

In this section, we present a new variational principle for purely dissipative evolution equations based on the
\emph{ansatz} that systems evolve in the direction of maximum entropy production (see
  Eq.~\eqref{Eq:MEPP_general_coupled} below)~\footnote{The entropy production can be expressed as the entropy rate of
  the system minus the entropy increase induced by heat flux exchange with the ambient
  space~\cite{prigogine1971structure}.}, so as to reach the equilibrium configuration as fast as possible. The
philosophy is therefore similar to SEA and MEPPs, yet different in its detailed formulation. In particular, the
proposed principle will provide a direct relation between the operator $K$ and physical constraints in the system, thus
shedding some light on the geometry of dissipative equations.
 %be shown to provide the structure of the operator $K$, and this will be related to physical constraints in the system. 

For simplicity, we first consider closed systems defined by a scalar variable and later generalize the obtained results to open systems and the vectorial setting. Illustrative examples are then chosen to 
demonstrate the applicability of the principle for both conserved and non-conserved fields, with explicit consideration of the boundary conditions. We note that non-conserved quantities do not have a flux associated to them, and therefore lie outside of the direct scope of Onsager's principle~\eqref{Eq:LeastDissipationEngergy}.

For a closed system out of equilibrium characterized by a scalar state variable $z$, the maximum entropy production ansatz is mathematically equivalent to the search of the velocity $\dot{z}$ maximizing $\dot S = \int \dot s\, d x = \langle DS, \dot z\rangle$, where $s$ is the entropy density and $S$ the total entropy of the system. The maximization is pointwise in the tangent space for fixed $z$, c.f.~Fig.~\ref{Fig:ConstrainedMaximization}. However, this problem is not well-posed unless the length of the vector $\dot{z}$ is prescribed, in which case the problem is reduced to the search of the optimal direction. This constraint is easily incorporated with a Lagrange multiplier, yielding a variational principle with Lagrangian
\begin{equation}
  \label{Eq:MaxEnt_L2}
 \mathcal{D}[\dot{z}] = \langle DS(z),\dot{z} \rangle -  \langle \dot{z},\eta(z) \dot{z}\rangle, 
\end{equation}
where the precise value of the length, which may depend on $z$, has been obviated since it does not participate in variations for fixed $z$. The evolution is then obtained by variations of~\eqref{Eq:MaxEnt_L2} with respect to $\dot z$, giving $\dot{z} = m(z)  DS(z)$, with $m(z)= \left(2\eta(z)\right)^{-1}\geq 0$, since entropy would decrease otherwise.  This shows that the $L^2_m$ gradient flow \eqref{Eq:L2_GF} with functional $S$ naturally results from the maximum entropy production principle in the absence of any physical constraint.

However, the evolution of $z$ is often subjected to conservation constraints of the form 
\begin{equation*}
\frac{d}{dt} \int_{\Omega} z \, dx = 0,
\end{equation*}
which naturally occurs when $z$ represents mass or energy. In this situation, the maximal dissipation occurs within the manifold of conserved $z$, 
\begin{equation*}
\dot{z}+ \nabla \cdot \J = 0,
\end{equation*}
where $\J=0$ on the boundary $\partial \Omega$ for a closed system. With an additional Lagrange multiplier $\lambda$, the variational principle at each point $z$ can then be written as
\begin{equation}
\label{Eq:MaxEnt_Wasserstein}
\mathcal{D}[\dot{z},\J,\lambda] = \langle DS,\dot{z} \rangle - \langle \lambda(z), \dot{z}+ \nabla \cdot \J \rangle - \langle \J, \mathbf{H}(z) \J\rangle,
\end{equation}
where the length constraint (measured with metric tensor $\mathbf{H}$) has now been placed on the unknown variable $\J$. We note that constraining the length of $\dot{z}$ as in \eqref{Eq:MaxEnt_L2} would leave $\J$ partially undetermined, and so would be the constitutive relations, such as Fourier's law for the case of heat conduction.

Variation with respect to $\J$  in \eqref{Eq:MaxEnt_Wasserstein} delivers 
\begin{equation*}
0= - \langle \lambda, \nabla \cdot \delta \J \rangle - \langle 2 \mathbf{H} \J, \delta \J \rangle \text{ for all }\delta \J,
\end{equation*}
which, after integration by parts, yields
$\nabla \lambda = 2 \mathbf{H} \J$.  Variations with respect to $\dot{z}$ and $\lambda$ give
\begin{equation*}
DS - \lambda = 0 \text { and } \dot{z} = - \nabla \cdot \J.
\end{equation*}
Altogether, this leads to a Wasserstein gradient flow  with functional $S$ and weight $\mathbf{M}=(2 \mathbf{H})^{-1}$ positive semi-definite, $\dot{z} = - \nabla \cdot \left(\mathbf{M} \nabla DS \right)$. The Wasserstein gradient flow \eqref{eq:wass} can be thus be understood as an $L^2$ gradient flow restricted to the manifold of conserved quantities. 

In general, systems are characterized by a set of state variables $\z$, some of which are conserved, $\z_c$, (e.g., energy, concentration), and some of which are not, $\z_u$, (e.g., interface position), i.e., $\z=[\z_u,\z_c]^T$. In this case the variational principle can be written as 
\begin{multline}
 \label{Eq:MEPP_general}
\mathcal{D}[\dot{\z},\J,\Lambda] =\langle DS,\dot{\z} \rangle - \langle \Lambda(\z), \dot{\z}_c+ \nabla \cdot \J \rangle 
- \langle \J, \mathbf{H}_c(\z) \J\rangle  \\
- \langle \dot{\z}_u,\mathbf{H}_u(\z) \dot{\z}_u\rangle,
\end{multline}
where now $\Lambda$ is a vectorial Lagrange multiplier, and $ \mathbf{H}_u$ and $ \mathbf{H}_c$ are second-order tensors. Similar derivations as above yield the evolution equations 
\begin{align*}
&\dot{\z}_u=(2\mathbf{H}_u)^{-1}\frac{\delta S}{\delta \z_u} = K_u \frac{\delta S}{\delta \z_u}, \\
&\dot{\z}_c=-\nabla \cdot \left( (2\mathbf{H}_c)^{-1}\nabla \frac{\delta S}{\delta \z_c} \right)= K_c \frac{\delta S}{\delta \z_c},
\end{align*}
which have an analogous structure to those previously obtained. However, for anisotropic materials, coupling between variables of different tensorial quantities is possible, and in this case, the Lagrangian shall be written as
\begin{multline}
 \label{Eq:MEPP_general_coupled}
\mathcal{D}[\dot{\z},\J,\Lambda] =\langle DS,\dot{\z} \rangle - \langle \Lambda(\z), \dot{\z}_c+ \nabla \cdot \J \rangle \\ 
- \langle [\dot{\mathbf{z}}_u,\mathbf{J}]^T, \mathbf{H}(\mathbf{z}) [\dot{\mathbf{z}}_u,\mathbf{J}]^T \rangle.
\end{multline}
Variations of this functional with respect to $\dot{\z}_u, \dot{\z}_c, \mathbf{J}$ and $\Lambda$ give
\begin{equation*}
2 \mathbf{H}
\begin{pmatrix}
\dot{\z}_u\\
\mathbf{J}
\end{pmatrix} =
\begin{pmatrix}
2 \mathbf{H}_u & 2 \mathbf{H}_{uc} \\
 2 \mathbf{H}^T_{uc} & 2 \mathbf{H}_c
\end{pmatrix}
\begin{pmatrix}
\dot{\z}_u\\
\mathbf{J}
\end{pmatrix}
= 
\begin{pmatrix}
\frac{\delta S}{\delta \z_u}\\
\nabla  \frac{\delta S}{\delta \z_c}
\end{pmatrix},
\end{equation*}
with $\dot{\z_c}=-\nabla \cdot \mathbf{J}$. Then, the evolution equations read
\begin{equation*}\label{Eq:EvolutionCoupled}
\begin{pmatrix}
\dot{\z}_u\\
\dot{\z}_c
\end{pmatrix}=
\begin{pmatrix}
\mathbf{M}_u & \mathbf{M}_{uc}\nabla \square \\
-\nabla \cdot \left(\mathbf{M}^T_{uc} \square \right) & -\nabla \cdot \left(\mathbf{M}_c \nabla\square  \right)
\end{pmatrix} 
\begin{pmatrix}
\frac{\delta S}{\delta \z_u}\\
\frac{\delta S}{\delta \z_c}
\end{pmatrix}=K DS,
\end{equation*}
where the symbol $\square$ indicates how the operator is applied to the vector $DS$. Further, $\mathbf{M}^{-1}=2 \mathbf{H}$, i.e.,
\begin{align*}
& \mathbf{M}_u= \frac{1}{2}\left[ \mathbf{H}_u-\mathbf{H}_{uc} \mathbf{H}_c^{-1}\mathbf{H}_{uc}^T \right]^{-1}, \\
& \mathbf{M}_c= \frac{1}{2}\left[ \mathbf{H}_c-\mathbf{H}_{uc}^T\mathbf{H}_u^{-1}\mathbf{H}_{uc} \right]^{-1},\\
& \mathbf{M}_{uc}= -\frac{1}{2}\left[ \mathbf{H}_u-\mathbf{H}_{uc} \mathbf{H}_c^{-1}\mathbf{H}^T_{uc} \right]^{-1}\mathbf{H}_{uc} \mathbf{H}_c^{-1}.
\end{align*}

This simple viewpoint of dissipative evolution equations via constrained maximization will be exemplified below for the equation of heat transfer and interface motion in open system, as blueprint for the derivation of other equations in a similar manner. 

%We remark that the variational approach presented here is analogous to that taken by Boltzmann to obtain the Boltzmann distribution $p_i = Z^{-1} \exp(- e_i / (k_B T))$ with $Z = \sum_i \exp(-e_i/(k_BT)$ as maximizer of the entropy in equilibrium among probability measures and average energy $\bar e$~\cite[Section 29]{Tolman},
%\begin{equation*}
%\sum_i k_B p_i \log p_i - \lambda \left[ \sum_i p_i e_i - \bar e  \right] -
%\mu\left[ \sum_i p_i -1\right] .
%\end{equation*}
%Then the Lagrange multiplier $\lambda$ takes the value $1/T$ as in the dynamic situation discussed in the next example. The method developed here thus extends in spirit Boltzmann's approach to dissipative evolutions.

\paragraph{Example: the heat equation and Fourier's law.} 
We now show that Fourier's law and the heat equation  follow directly from the postulate of maximum entropy production. For an open system, the Lagrangian of the maximum entropy production principle is constructed by subtracting the entropy flow entering the boundary of the domain from the total entropy rate. Then the entropy increase considered exclusively originates from the internal production, in accordance with the second law of thermodynamics. Assuming that the system is completely characterized by the internal energy, and taking also the conservation of energy into account, the Lagrangian reads
\begin{multline}\label{Eq:VariationalHeat}
\mathcal{D}[\dot{e},\lambda,\q]= \int_{\Omega} \frac{\partial s}{\partial e}\dot{e} \, dx+ \int_{\partial \Omega} \frac{\q\cdot \mathbf{n}}{T} \, dx \\ - \int_{\Omega} \lambda \left(\dot{e} + \nabla \cdot \q \right)\, dx
- \int_{\Omega} \q^T \mathbf{H} \q \, dx,
\end{multline}
where $\mathbf{n}$ is the outer normal to the domain, $\q$ is the heat flux, and $s$ and $e$ represent the entropy and energy per unit volume, respectively. From basic thermodynamic relations, assuming local thermodynamic equilibrium, $\dot{s} = \frac{1}{T}\dot{e}$. Therefore, variations with respect to $\dot{e}$, $\lambda$ and $\q$, assuming boundary conditions in $T$ (boundary conditions in $q$ would imply $\delta \q = 0$ on $\partial \Omega$, and lead to the same evolution equation) yield
\begin{align*}
&\frac{1}{T}- \lambda =0, \\
& \dot{e} + \nabla \cdot \q = 0, \\
&\int_{\Omega} \Big[ \nabla \cdot\left(\frac{\delta \q}{T} \right)-\lambda \nabla \cdot (\delta \q) -2 \mathbf{H} \q \cdot \delta \q \Big] \, dx = 0 \qquad  \forall \delta \q,
\end{align*}
which combined give the equation of heat transfer, with $\mathbf{K} \coloneqq (2 \mathbf{H} T^2)^{-1}$,
\begin{equation}
 \label{eq:heat}
\frac{\partial e}{\partial T} \dot{T} = \dot{e} = -\nabla \cdot \left( (2\mathbf{H})^{-1}\nabla\left(\frac{1}{T} \right)\right) = \nabla\cdot \left( \mathbf{K} \nabla T \right).
\end{equation}

We remark that the analogous derivation via Onsager's principle of least dissipation, i.e.,
\begin{equation}
  \label{Eq:OnsagerEnergy}
  \mathcal{D}[\X] = \q\cdot \X - \frac{1}{2} \X \mathbf{L} \X,\quad   \text{with  }\X= - \nabla   \left(\frac{1}{T} \right)
\end{equation}
leads to Fourier's law $\q = \mathbf{L}\X$, which, complemented with the first law of thermodynamics, yields~\eqref{eq:heat} with $\mathbf{L}=(2\mathbf{H})^{-1}$. However, the physical motivation of the Lagrangian in Eq.~\eqref{Eq:VariationalHeat} seems more natural than that of \eqref{Eq:OnsagerEnergy}.

%\begin{figure}[t]
%\begin{center}
%    {\includegraphics[width=0.4\textwidth]{ConstrainedMaximization.pdf}}
%    \caption{
%    Evolution of a closed system at $z(t)$ with and without the presence of a constraint. The equilibrium configuration coincides with the maximum of the entropy functional $S(z)$.}
%    \label{Fig:ConstrainedMaximization}
%\end{center}
%\end{figure}

\paragraph{Example: Interface motion in an isotropic medium.}
Next, we consider a two-phase system separated by an interface, which we characterize by an additional variable $\phi$ in the spirit of a phase field model \citep{provatas2011phase}. Following a similar strategy as in the previous case, the evolution of the interface coupled to the heat equation can be obtained as the extremum of
\begin{multline*}
\mathcal{D}[\dot{e},\dot{\phi},\lambda,\q]= \int_{\Omega}\dot{s}\, dx+ \int_{\partial \Omega} \frac{\q\cdot \mathbf{n}}{T} \, dx \\
 - \int_{\Omega} \lambda \left(\dot{e} + \nabla \cdot \q \right)\, dx- \int_{\Omega} \q\cdot \mu \q \, dx - \int_{\Omega} \dot \phi \eta \dot \phi \, dx.
\end{multline*}
Assuming the existence of a thermodynamic relation for the energy density $e$ of the form $e = e(s,\phi, \nabla \phi)$,
\begin{equation*}
de = Tds + \left(\frac{\partial e}{\partial \phi} \right)_{s,\nabla \phi} d\phi +\left(\frac{\partial e}{\partial \nabla \phi} \right)_{s,\phi} d\nabla\phi ,
\end{equation*}
where subscripts indicate the variables that are held fixed. Its Legendre transform with respect to the entropy density $s$ is the Helmholtz free energy $f$,
\begin{equation*}
df = -sdT + \left(\frac{\partial e}{\partial \phi} \right)_{s,\nabla \phi} d\phi +\left(\frac{\partial e}{\partial \nabla \phi} \right)_{s,\phi} d\nabla\phi .
\end{equation*}
 One then obtains
\begin{align*}
\dot{s} &= \frac{1}{T}\dot{e} - \frac{1}{T}\left(\frac{\partial e}{\partial \phi} \right)_{s,\nabla \phi} \dot{\phi }- \frac{1}{T}\left(\frac{\partial e}{\partial \nabla \phi} \right)_{s,\phi} \nabla \dot{\phi} \\
&=\frac{1}{T}\dot{e} - \frac{1}{T}\left(\frac{\partial f}{\partial \phi} \right)_{T,\nabla \phi} \dot{\phi }- \frac{1}{T}  \nabla \dot{\phi}.
\end{align*} 
% One then obtains, with $f$ being the Helmholtz free energy,
%\begin{align*}
%\dot{s} &=\frac{1}{T}\dot{e} - \frac{1}{T}\left(\frac{\partial f}{\partial \phi} \right)_{T,\nabla \phi} \dot{\phi }- \frac{1}{T}%\left(\frac{\partial f}{\partial \nabla \phi} \right)_{T,\phi} \nabla \dot{\phi}  .
%\end{align*}
As in the previous example, we obtain the heat equation from variations of $\mathcal{D}$ with respect to $\dot{e}, \lambda$ and $q$, while variations with respect to $\dot \phi$ yield the evolution of the interface, 
\begin{align*}
&\dot{e} = \nabla \cdot \left( k \nabla T \right), \ \text{with }k \coloneqq (2\mu T^2)^{-1}\\
&2 \eta \dot{\phi} = - \frac{\delta \int f/T \, d x }{\delta \phi}
= - \frac{\partial (f/T)}{\partial \phi} + \nabla \cdot \frac{\partial (f/T)}{\partial \nabla \phi}.
\end{align*}
Thus the interface is driven by the Massieu potential $- f/T$, whose relevance has been noted in SEA \cite{Beretta2006, Beretta2007, Beretta2009}, in the GENERIC setting~\cite{mielke2011formulation} as well as in large deviation theory~\cite{Touchette2009a}. We note that the derivation of this evolution in the Onsager formalism is nontrivial as $\phi$ does not have a flux and a corresponding thermodynamic force.

\section{Stochastic evolution and large deviations}
\label{sec:SDELDP}

In this section, we show that the proposed variational formulation for purely dissipative equations based on physical considerations is further supported by a large deviation principle (LDP) associated to stochastically
perturbed gradient flows. The LDP provides the probability of a given evolution to occur, and therefore intrinsically
contains a variational principle for the most likely path. Large deviation arguments have recently been used to connect particle models to gradient flows, for example in~\cite{Adams:09a,Adams:12a,Mielke:14a}, and have also led to variational formulations of systems in GENERIC form~\cite{Duong:13a}. 

Specifically, let $\z=\z(t,\ux)$ be a vector field that evolves in $t\in [0,T]$ according to a stochastic gradient
flow with small noise,
\begin{equation}\label{Eq:StochasticGF}
d\z = K(\z)DS(\z) dt + \sigma(\z)\sqrt{\epsilon}\ d\W_{t,x},
\end{equation}
where $\W_{t,x}$ is a vector of independent Brownian sheets, i.e., $\mathbb{E}[ \Ws_{i;t,x} \Ws_{j;s,y} ] = \delta_{ij}\delta(t-s)\delta(\ux-\uy)$, with $\delta_{ij}$ the Kronecker delta function and $\delta(\mathbf{x})$ the Dirac delta function. Further, 
$\sigma(\z)$ is an operator acting on $d\W_{t,x}$, and $\epsilon$ is a small parameter controlling the strength of the noise. The stochastic calculus is to be understood in the It\^o sense. 

The probability distribution for $\z(t,\mathbf{x})$ satisfying \eqref{Eq:StochasticGF} may be obtained from that of simpler processes using the theory of large deviations and the contraction principle \cite{Freidlin1984a}. Indeed, by Schilder's theorem, the probability distribution of the solutions to the vectorial ordinary differential equation $d\uu=\sqrt{\epsilon} d\W_t$, with $\W_t$ a vector of time white noises,  $\mathbb{E}[\Ws_i(t) \Ws_j(s)] = \delta_{ij} \delta(t-s)$, follows 
\begin{equation} \label{Eq:LDP_Bt}
\mathbb{P}[\uu(t)\approx\uvarphi(t)]\propto e^{-\frac{1}{\epsilon}I[\uvarphi]}, \text{ where } I[\uvarphi]=\frac1 2 \int_{0}^{T}|\dot{\uvarphi}|^2 \, dt
\end{equation}
is called the rate functional. In words, the probability for $\uu(t)$ undergoes an exponential decay with rate $1/\epsilon$, and narrows as $\epsilon \rightarrow 0$ around the deterministic solution $\dot{\uvarphi}=0$. Then, the probability distribution for $\uv(t,\ux)$ satisfying $d\uv=\sqrt{\epsilon}d\W_{t,x}$ can be obtained by expanding $\uv(t,\ux)$ and $\W_{t,x}$ with orthonormal  basis functions $e_k(\ux)$ for the domain \cite{faris1982large,Freidlin1988a},
\begin{equation}
\uv= \sum_k \mathbf{A}^k_t e_k,\quad \W_{t,x} = \sum_k \mathbf{B}^k_t e_k,
\end{equation} 
where $\mathbf{B}^k_t$ are independent Brownian motions (direct computations show that $\mathbb{E} [ \Ws_{i;t,x} \Ws_{j;s,y} ] = \delta_{ij}\delta(t-s)\delta(\ux-\uy)$). The partial differential equation $d\uv=\sqrt{\epsilon}d\W_{t,x}$ is then equivalent to the system of vectorial ordinary differential equations $d\uA^k = \sqrt{\epsilon} d\W^k_t$; and the rate functional of the associated large deviation principles, for  $\uvarphi = \sum_k \mathbf{C}^k e_k$ (see, e.g., \cite{faris1982large,Freidlin1988a}), can be readily obtained from \eqref{Eq:LDP_Bt}
\begin{equation}
I[\uvarphi ]= \frac 1 2\int_0^T \sum_k |\dot{\mathbf{C}}^k|^2 \, dt =\frac 1 2\int_0^T\|\dot{\uvarphi}\|^2 \, dt.
\end{equation}

The solutions to \eqref{Eq:StochasticGF} can be seen as $\z(t,\mathbf{x})= M \W_{t,x}= M(\uv/\sqrt{\epsilon})$, where $M$ is an operator. If $M$ is continuous (see \cite{budhiraja2008large} for measurable functions), then, by the contraction principle, $\z$ follows a large deviation principle \cite{Sowers1992a} with functional  $I[\boldsymbol \varphi ]=\frac 1 2\int_0^T\|\dot{M^{-1} \boldsymbol \varphi}\|^2 \, dt$, i.e., 
\begin{multline} \label{Eq:LDP}
\mathbb{P}[\z (t,\ux)\approx \boldsymbol\varphi(t,\ux)] \propto  \\
\exp\left(-\frac 1 \epsilon \frac{1}{2} \int_{0}^{T} \| \dot{\boldsymbol\varphi} - K(\varphi)\ DS(\boldsymbol \varphi) \|^2_{\left( \sigma  \sigma^* \right)^{-1}} \, dt \right),
\end{multline}
assuming $\left(   \sigma\sigma^*  \right)^{-1}$ defines a norm, with $\sigma^*$ being the adjoint operator of $\sigma$. This result follows the spirit of Onsager and Machlup~\cite{onsager1953fluctuations}, for general gradient flow structures; however, the probability distribution obtained is not a function of the thermodynamic forces and fluxes as in the original formulation by Onsager, but of the variable $\z$ and $\dot{\z}$. This difference is analogous to that of~\eqref{Eq:VariationalHeat} and~\eqref{Eq:OnsagerEnergy}.

\section{Maximum entropy production from large deviations}
Equation~\eqref{Eq:LDP} shows that the most likely path is the one that maximizes the exponent and thus minimizes $\int_{0}^{T} \| \dot{\z} - K(\z)\ DS(\z) \|^2_{(\sigma \sigma^*)^{-1}} \, dt$. This minimum is attained by pointwise optimization (over $\dot{\z}$ for fixed $\z$ at every instant of time), giving
\begin{equation}\label{Eq:integrand}
\min_{\dot{\z}} \| \dot{\z} - K(\z)\ DS(\z) \|^2_{(\sigma\sigma^* )^{-1}}\ .
\end{equation}

Equation \eqref{Eq:integrand} represents a variational principle for the deterministic gradient flow, which, for $K \propto \sigma \sigma^*$, is shown below to be equivalent to Eq.~\eqref{Eq:MaxEnt_L2} for $L^2$ gradient flows, to Eq.~\eqref{Eq:MaxEnt_Wasserstein} for the Wasserstein evolution, and to Eqs.~\eqref{Eq:MEPP_general} and \eqref{Eq:MEPP_general_coupled} for the combined vectorial case. Indeed, expanding the squares in Eq.~\eqref{Eq:integrand} yields the variational problem
\begin{equation}\label{Eq:D_Machlup_form}
\max_{\dot{\z}} \Big[ \langle DS(\z), \dot{\z}\rangle - \Phi(\dot{\z}) - \Psi(\z) \Big] 
\end{equation}
with $ \Phi(\dot{\z}) = \frac{1}{2} \| \dot{\z} \|^2_{K^{-1}} $ and $\Psi(\z) = \frac{1}{2} \| DS(\z) \|^2_{K} $, where the latter does not affect the optimal evolution. One has $\Phi \neq \Psi$ in the presence of fluctuations, whereas for the optimal path $\dot{S} = 2 \Phi = 2 \Psi$ holds.  

The Lagrangian for the $L^2_m$ gradient flow ($K = L^2_m$), 
Eq.~\eqref{Eq:MaxEnt_L2}, can be rewritten in the form of \eqref{Eq:D_Machlup_form},
\begin{equation*}
\mathcal{D} =\langle DS(z), \dot{z}\rangle -  \frac {1}{ 2} \|\dot{z} \|^2_{L^2_m} =\langle DS(z), \dot{z}\rangle  - \Phi(\dot{z}),
\end{equation*}
with $\eta= \frac{1}{2m}$, and $\| \|_{L^2_m}$ as defined in the Appendix. 
An equivalent result is obtained for the Wasserstein gradient flow ($W_{\mathbf{M}}=K^{-1}$), noting that the last term of Eq.~\eqref{Eq:MaxEnt_Wasserstein}, with $\mathbf{H}=\frac{1}{2}\mathbf{M}^{-1}$ and $\dot{z}+\nabla \cdot \J=0$, can be rewritten as
\begin{equation}\label{Eq:Phi_W}
\langle \J, \mathbf{H}(z) \J\rangle = \frac{1}{2}\langle \J,\J\rangle_{L^2_{\mathbf{M}^{-1}}} = \frac{1}{2}\langle \dot{z}, \dot{z}\rangle_{W_\mathbf{M}}=\Phi(\dot{z}).
\end{equation}
The vectorial $L^2$ norm is defined analogously to the scalar case, and the second equality in \eqref{Eq:Phi_W} is detailed in the Appendix. Similarly derivations for the vectorial case considered in \eqref{Eq:MEPP_general} lead to
\begin{multline}
\langle \J, \mathbf{H}_c(\z) \J\rangle  + \langle \dot{\z}_u,\mathbf{H}_u(\z) \dot{\z}_u\rangle = \| \z_u\|^2_{K_u^{-1}} + \| \z_c\|^2_{K_c^{-1}} \\
= \|\z \|^2_{K^{-1}}=\Phi(\dot{z}),
\end{multline}
with $K^{-1}= \text{diag}\left(K_u^{-1},K_c^{-1} \right)$. For the coupled case considered in Eq.~\eqref{Eq:MEPP_general_coupled}, $K$ is a full matrix and its inverse reads
\begin{equation}
K^{-1}=\begin{pmatrix}
2\mathbf{H}_u & -2\mathbf{H}_{uc} \left(\nabla \cdot \right)^{-1} \\
\nabla^{-1} 2 \mathbf{H}^T_{uc} & -\nabla^{-1} 2 \mathbf{H}_{c} \left(\nabla \cdot \right)^{-1}
\end{pmatrix},
\end{equation}
where the inverted divergence $ \left(\nabla \cdot \right)^{-1} $ and inverted gradient $\nabla^{-1}$ are to be interpreted in appropriate spaces. The relations $KK^{-1}=K^{-1}K=I$ immediately follow from $2\mathbf{M}\mathbf{H}=2\mathbf{H}\mathbf{M}=\mathbf{I}$. Then, one similarly obtains that
\begin{equation}
\left\langle \begin{pmatrix}
\dot{\z}_u \\
\mathbf{J}
\end{pmatrix},
\begin{pmatrix}
 \mathbf{H}_u & \mathbf{H}_{uc} \\
  \mathbf{H}^T_{uc} &  \mathbf{H}_c
\end{pmatrix}
 \begin{pmatrix}
\dot{\z}_u \\
\mathbf{J}
\end{pmatrix}
\right \rangle = \frac{1}{2}\|\dot{\z}\|^2_{K^{-1}}=\Phi(\dot{\z}).
\end{equation}

The variational principles of Eqs.~\eqref{Eq:MaxEnt_L2}--\eqref{Eq:MEPP_general_coupled} can therefore be written as 
$\min_{\dot{\z}} \| \dot{\z} - K(\z)\ DS(\z) \|^2_{K^{-1}}$.
 %This is also the case for the vectorial case considered in \eqref{Eq:MEPP_general}, whose derivation follows in a very similar manner (note $\|\z \|^2_{K^{-1}}=\| \z_u\|^2_{K_u^{-1}}+\| \z_c\|^2_{K_c^{-1}}$).
We thus observe that the diagram of Fig.~\ref{Fig:Diagram} commutes if $K\propto\sigma \sigma^*$, which represents a fluctuation-dissipation relation in infinite dimensions. 

\textbf{Square root of the Wasserstein operator.}

We now discuss the expression $\sigma=\sqrt{K}$ encountered in the fluctuation-dissipation statement above for the Wasserstein operator. In general, for a given positive semi-definite self-adjoint $K$ there are several choices $\sigma_1 \sigma_1^* = \sigma_2 \sigma_2^* = K$. However, only $\sigma \sigma^*$ appears in 
the generator and thus the solutions to the corresponding Fokker-Planck equations for different roots are statistically equivalent~\cite{ottinger1996stochastic}. For Wasserstein gradient flows we only consider $\sigma=\sqrt{K}$ of divergence form, to have a conservative noise, i.e., $\sigma d\W = \nabla \cdot \mathbf{j}$, where  $\epsilon=1$ for simplicity. Then, for the  Wasserstein metric with mobility $\mathbf{M}$
\begin{multline}
 \Big \langle d\W_{t,x}, 
d\W_{t,x} \Big \rangle_{L^2} = \Big \langle \sigma d\W_{t,x}, \sigma d\W_{t,x} \Big \rangle_{W_{\mathbf{M}}} =\Big \langle  \mathbf{j}, \mathbf{j} \Big \rangle_{L^2_{\mathbf{M}^{-1}}} \\
=\Big \langle \mathbf{M}^{-1/2} \mathbf{j},\mathbf{M}^{-1/2} \mathbf{j} \Big \rangle_{L^2}, 
\end{multline}
from which one obtains that $d\W_{t,x} = \mathbf{M}^{-1/2}\mathbf{j}$, or equivalently, 
$\sigma d\W_{t,x} = \nabla \cdot \left(\mathbf{M}^{1/2} d\W_{t,x} \right)$. For the diffusion equation \eqref{Eq:DiffusionEquation} with unit diffusion constant, the stochastic version given by \eqref{Eq:StochasticGF} with $\sigma=\sqrt{K}$ reads
\begin{equation}
\dot \rho = \Delta \rho +  \nabla \cdot \left( \rho^{1/2} d\W_{t,x} \right).
\end{equation}
This equation of fluctuating hydrodynamics~\cite{eyink1990dissipation} is known as Dean-Kawasaki model~\cite{dean1996langevin,   kawasaki1998microscopic,chavanis2011brownian}.

\section{Conclusions}

We provide two independent derivations of a variational principle governing dissipative evolution equations of the form $\dot{\z}=K(\z) DS(\z)$. The first is based on the maximization of the entropy production within the manifold of constraints, extending Onsager's original approach, and provides insight into the geometry of the gradient flow structure ($K$). In particular, the principle captures multiple metrics: one which is related to a thermodynamic length, and others that may result from the constraints in the system, such as conservation of mass or energy. The first metric is here taken as the $L^2$ metric and is in principle unknown (an extension to general metrics, as in SEA, is yet to be explored), whereas the second one is an outcome of the variational statement. By means of this procedure, the Wasserstein metric is here shown to be equivalent to the constrained $L^2$ metric associated to conserved fields. 
The second approach for obtaining the variational statement is based on the large deviation principle for the gradient flows augmented by a noise term $\sigma(\z)\sqrt{\epsilon}\ d\mathbf{B}_{t,x}$, and is shown to be equivalent to the previously derived principle for $K\propto\sigma \sigma^*$. This represents a fluctuation-dissipation relation in infinite dimensions and endows the exponent of the large deviation principle with the usual interpretation of an entropy (dissipation) shortfall between a given path and the optimal one \cite{Varadhan2010}. 

\section{Appendix}

We write the weighted $L^2$ norm as $\langle v,w \rangle_{L^2_{m}} = \int m(\mathbf{x}) v(\mathbf{x}) w(\mathbf{x})\, dx$, and denote $L^2\coloneqq L^2_1$ and $\langle, \rangle \coloneqq \langle, \rangle_{L^2}$ (note that for square integrable functions, $\langle, \rangle_{L^2}$ is equivalent to the duality pairing). Then the weak formulation of the $L^2_{m}$ gradient flow for the diffusion equation is ($z \coloneqq \rho$  and $S=S_1$ in Eq.~\eqref{Eq:L2_GF})
\begin{align*}
\langle \dot \rho,v \rangle_{L^2} &= \langle DS_1(\rho),v \rangle_{L^2_{m}}  \\ &= - \langle \nabla\rho ,\nabla v \rangle_{L^2_{m}}  = \langle \nabla \cdot (m(\rho) \nabla\rho) , v \rangle_{L^2}  . 
\end{align*}

For the Wasserstein gradient flow, if $\dot z_i = W_{\mathbf{M}}^{-1}p_i = - \nabla \cdot \left( \mathbf{M} \nabla p_i\right) = \nabla \cdot \J_i$ with $\nabla p_i = 0$ on $\partial \Omega$, the Wasserstein norm is
\begin{align*}
  \langle \dot z_1, \dot z_2 \rangle_{W_{\mathbf{M}}} &\coloneqq 
  \langle \mathbf{M} \nabla p_1, \nabla p_2 \rangle_{L^2} 
  = - \langle \nabla \cdot  \mathbf{M} \nabla p_1,  p_2 \rangle_{L^2} \\
  &=  \langle \dot z_1, W_{\mathbf{M}} \dot z_2 \rangle_{L^2}.
\end{align*}
The second expression is known as the $H_1$ seminorm with weight $\mathbf{M}$, 
$\langle p_1, p_2\rangle_{\HS{M}} \coloneqq \langle \mathbf{M} \nabla p_1, \nabla p_2 \rangle_{L^2}$. We write (see~\cite[Appendix D]{FengKurtz} for details)
\begin{multline*}
   \langle \dot z_1, \dot z_2 \rangle_{W_{\mathbf{M}}} =  \langle \mathbf{M} \nabla p_1, \nabla p_2 \rangle_{L^2} = \langle \mathbf{M}^{-1} \J_1,\J_2 \rangle_{L^2} \\
  \eqqcolon \langle \dot z_1 , \dot z_2 \rangle_{\HSd{\mathbf{M}}}.
\end{multline*}
With this notation, it is straightforward to calculate the weak formulation of the diffusion equation as a Wasserstein gradient flow ($z \coloneqq \rho$, $\mathbf{M}_2=m\rho\mathbf{I}=m_2\mathbf{I}$  and $S=S_2$ in Eq.~\eqref{eq:wass}), 
\begin{equation*}
  \langle \dot \rho,\dot z_2\rangle_{W_{\mathbf{M}_2}} = 
\langle \nabla DS_2(\rho), m_2 \nabla p_2(z_2) \rangle_{L^2}. 
\end{equation*}

{\bf Acknowledgments.} The authors thank M.~von Renesse, P.~Ayyaswamy, D.~Kelly, R.~Jack, V.~Maroulas, M.~Renger and E.~Vanden-Eijnden for valuable comments. This work was partially supported by the UK's Engineering and Physical Sciences Research Council Grant EP/K027743/1 (to JZ), the Leverhulme Trust (RPG-2013-261) and GW4 grants GW4-IF2-026 and GW4-AF-005. We appreciate helpful suggestions from the reviewers.

%\bibliographystyle{apsrev4-1}
%\bibliography{Biblio}


\end{document}